\documentstyle[preprint,aps]{revtex}

 \newcommand \be {\begin{equation}}
\newcommand \bea {\begin{eqnarray} \nonumber }
\newcommand \ee {\end{equation}}
\newcommand \eea {\end{eqnarray}}
 
 \newcommand \bi {\bibitem}
\newcommand \s {\sigma}

\newcommand \De {\Delta}

 \newcommand \al {\alpha}

\begin{document}
\draft
\preprint{MA/UC3M/09/96}
\title{Relaxational behavior of the infinite-range Ising spin-glass
in a transverse field}

\author{J. V. Alvarez and Felix Ritort}
\address{Departamento de Matem{\'a}ticas,\\
Universidad Carlos III, Butarque 15\\
Legan{\'e}s 28911, Madrid (Spain)\\
E-Mail: vicente@dulcinea.uc3m.es\\
E-Mail: ritort@dulcinea.uc3m.es}

\date{\today}
\maketitle

\begin{abstract}
We study the zero-temperature behavior of the infinite-ranged Ising spin
glass in a transverse field. Using spin summation techniques and Monte
Carlo methods we characterize the zero-temperature quantum
transition. Our results are well compatible with a value
$\nu=\frac{1}{4}$ for the correlation length exponent, $z=4$ for the
dynamical exponent and an algebraic decay $t^{-1}$ for the
imaginary-time correlation function. The zero-temperature relaxation of
the energy in the presence of the transverse field shows that the system
monotonically reaches the ground state energy due to tunneling processes
and displays strong glassy effects.
\end{abstract} 

\vfill
\pacs{64.70.Nr, 64.60.Cn}

\vfill

\narrowtext
\section{Introduction}

The purpose of this work is to present some results concerning the
zero-temperature critical and relaxational behavior of the Ising spin
glass in the presence of a transverse magnetic field. While classical
spin-glasses have been extensively studied during the recent years, the
role of the quantum effects in the low-temperature regime are not so
well understood. In particular, a large amount of work has been devoted
to the study of the one-dimensional case \cite{Fisher} and the
mean-field theory \cite{BrayMoore} \cite{GoLa}.  These two limiting cases
seem to capture one of the most relevant features associated to the
quantum fluctuations, i.e. the presence of tunneling effects at
zero-temperature. The effect of the transverse field is to allow the
sytem to jump over the free energy barriers even at zero temperature.
In this work we will focus our attention in the study of the
zero-temperature critical behavior and on the nature of the relaxational
dynamics. We have considered the infinite-range model where some
analytical results can be obtained. The infinite-ranged model has been
studied in several works. In particular, the phase diagram of the model
has been obtained using spin summation techniques \cite{GoLa,ThirLiKirk}
while Miller and Huse \cite{MiHu} have obtained the imaginary-time
correlation function at the zero temperature critical point using a
theoretical analysis. On the other hand recent numerical work
\cite{RiYo,GBHu} reveals that the Monte Carlo method can yield good
estimates of the critical exponents associated to the quantum transition
using finite-size scaling techniques.

Our purpose is two fold. First we want to show how the Monte Carlo
technique used in \cite{RiYo,GBHu} is a powerful tool in order to
determine the critical point and the critical exponents in the
mean-field case. This will be done comparing the results obtained using
finite-size scaling and numerical spin summation methods. Once the
critical field is obtained we will obtain the main critical exponents
$z$ and $\nu$ and we will study the decay of the imaginary-time
correlation function at the critical point. Unfortunately, our results
are in disagreement with the theoretical prediction of Miller and Huse
\cite{MiHu}. Second, we will consider the role of the quantum
fluctuations on the zero-temperature relaxational behavior of the
model. While these last results concern the quantum infinite-ranged
model we expect that our main conclusions are valid also in the
short-ranged case.

\section{The Model}

The model we are interested is defined by the Hamiltonian,

\be
{\cal H}=-\sum_{i<j}\,J_{ij}\s_i^z\,\s_j^z-\Gamma\sum_i\s_i^x
\label{eq1}
\ee

where the $\lbrace\s_i; i=1,N\rbrace$ are the Pauli spin matrices and
$\Gamma$ is the transverse field. The $J_{ij}$ are Gaussian distributed
variables with zero mean and variance $\frac{1}{N}$. For $\Gamma=0$ the
model reduces to the classical Sherrington-Kirkpatrick spin-glass model
\cite{ShKi}. It is well known \cite{Tr} \cite{Su} that the ground state
energy of the above Hamiltonian can be written as the free energy of a
classical model with an extra imaginary-time dimension in the following
way,

\be E_g(\Gamma)=-\lim_{\beta\to\infty}\lim_{M\to
\infty}\frac{\log(Z_{eff})}{N\beta}
\label{eq2}
\ee

\noindent 
where 
\be
Z_{eff}=Tr_{\s_i}exp(-\beta H_{eff}(\Gamma,M,\beta))=\sum_{\s_i=\pm 1}
\,exp(A\sum_{i<j}\sum_{t=1}^M\,J_{ij}\s_i^t\,\s_j^t\,+\,
B\sum_{i=1}^N\sum_{t=1}^M\s_i^t\s_i^{t+1}+C)
\label{eq3}
\ee

\noindent
and the spins $\s_i$ are classical variables which can take the
values $\pm 1$. The parameters $A$ and $B$ are given by,

\begin{eqnarray}
A=\frac{\beta}{M}\nonumber\\
B=\frac{1}{2}\log(coth(\frac{\beta\Gamma}{M}))\\
\label{eq3b}
C=\frac{MN}{2}\log(\frac{1}{2} sinh(\frac{2\beta\Gamma}{M}))\nonumber
\end{eqnarray}

In the limit $M\to\infty$ the parameters $A$ and $B$ are highly
anisotropic (the coefficient $A$ goes to zero while $B$ goes to
infinity).  This makes extremely difficult to perform Monte Carlo
numerical tests of the quantum model. It has been recently shown
\cite{RiYo,GBHu} that it is better to work with a different Hamiltonian
which nevertheless lies in the same universality class.  To this end, we
have considered the family of models with parameters $A=\beta_{cl}$,
$B=\beta_{cl}^n$ , $C=0$. Within this family of models the parameter
$\beta_{cl}$ has the role of the inverse of a classical temperature (not
to be confused with the real temperature) which controls the intensity
of the quantum fluctuations. In some sense, this effective classical
temperature $\frac{1}{\beta_{cl}}$ plays the role of a transverse field
in the {\em true} model (\ref{eq3b}).  In this way, the new Hamiltonian
is a more isotropic one. Also, in case universality holds, we expect the
mean-field critical exponents to be independent of the particular model
considered. We have concentrated our attention in the previous models
with $n=1$ (model (a)), and $n=2$ (model (b)) and we have studied them
using the Monte Carlo method and spin summation techniques. While our
Monte Carlo numerical results are consistent with the universality
hypothesis we have discovered that model (a) is still hampered by strong
Monte Carlo sampling problems while model (b) gives more confident
results.

\section{Spin Summation Results} 

In order to apply the spin summation techniques we have analitically
solved the previous model eq.(\ref{eq3}) using the replica trick with
general coefficients $A$ and $B$. The analytical solution of the
infinite-range model has been already considered in the literature
\cite{BrayMoore,ThirLiKirk} and here we will only remind the results. Applying the replica
trick and performing the usual technical steps in the theory of spin
glasses i.e. (introducing the order parameters and decoupling the different
sites) one gets the effective free energy,

\be
F_{cl}=-\frac{\overline{\log(Z_{eff})}}{N\beta_{cl}}=
\lim_{n\to 0}\frac{\overline{Z_{eff}^n}}{Nn\beta_{cl}}=
\lim_{n\to 0}\frac{A[Q,R]}{n\beta_{cl}}
\label{eq4}
\ee

where $\overline{(..)}$ stands for average over the disorder 
and $n$ is an integer which denotes the number of replicas. The 
saddle-point free energy $A[Q,R]$ reads,

\be
A[Q,R]=\frac{A^2}{4}\,(\sum_{\al\ne\beta}\sum_{t,t'} 
(Q_{\al\beta}^{t\,t'})^2+\sum_{\al}\sum_{t\ne t'}(R_{\al}^{t\,t'})^2)
-\log F[Q,R]
\label{eq5} 
\ee

with

\be
F[Q,R]=\sum_{\s_{\al}^t}\exp[B\sum_{t,\al}\s_{\al}^{t}\s_{\al}^{t+1}\,+\,
\frac{A^2}{2}(\sum_{\al\ne\beta}\sum_{t,t'}Q_{\al\beta}^{t\,t'}
\s_{\al}^t\s_{\beta}^{t'}\,+\,\sum_{\al}\sum_{t\ne t'}R_{\al}^{t\,t'}
\s_{\al}^t\s_{\al}^{t'})]
\label{eq6} 
\ee

The indices $\al,\beta=1,..,n$ stand for replica indices while the 
indices $t,t'=1,..,M$ run over the imaginary-time direction with periodic
boundary conditions (i.e. $\s_{\alpha}^{M+1}=\s_{\alpha}^{1}$). The saddle point equations 
yield the order parameters $Q$ and $R$,

\be
Q_{\al\beta}^{t\,t'}=\langle \s_{\al}^t\s_{\beta}^{t'}\rangle~~~~~~~
R_{\al}^{t\,t'}=\langle \s_{\al}^t\s_{\al}^{t'}\rangle\\
\label{eq7}
\ee

where the thermal averages $\langle...\rangle$ are done over the effective
partition function defined in eq.(\ref{eq6}).  
To solve the previous equation we impose the static condition 
(i.e. no dependence on the imaginary-time variables $t,t'$) in the set of 
parameters $Q$ while the $R's$ are assumed to be no static but 
traslationally-time invariant, i.e. depend only on the difference of times 
$t-t'$ . In order to determine the critical value of $\beta_{cl}$
it is enough to consider replica symmetry. In this case the order parameters
assume the form $Q_{\al\beta}^{t\,t'}=q, R_{\al}^{t\,t'}=R(t-t')$ an 
the free energy reads,

\be
\beta f=\frac{A^2}{4}\sum_{t\ne t'}(R^{t\,t'})^2\,-\,\frac{A^2M}{4}(1-2q)
\,-\,\frac{A^2M}{2}(1-q)\,-\,\int_{-\infty}^{\infty}
\frac{dx}{(2\pi)^{\frac{1}{2}}}e^{-\frac{x^2}{2}}\log\Theta(x)
\label{eq8}
\ee

where the function $\Theta(x)$ is given by,

\be
\Theta(x)=\sum_{\s^t}\exp(\Xi(x,\s))=\sum_{\s^t}\exp\Bigl 
(B\sum_t\s^t\s^{t+1}\,+\,
(A^2q)^\frac{1}{2}x\sum_t\s^t\,+\,\sum_{t\ne t'} (R_{tt'}-q)\s^t\s^{t'}
\Bigr )
\label{eq9}
\ee

and the order parameters $q$ and $R(t-t')$ can be obtained solving the 
equations,

\be
q=\int_{-\infty}^{\infty}\frac{dx}{(2\pi)^{\frac{1}{2}}}e^{-\frac{x^2}{2}}
\Bigl ( \frac{\sum_{\s}\s^t\exp(\Xi(x,\s))}{\Theta(x)}\Bigr )^2 \\
\ee
\be
R(t-t')=\int_{-\infty}^{\infty}\frac{dx}{(2\pi)^{\frac{1}{2}}}e^{-\frac{x^2}{2}}
\Bigl ( \frac{\sum_{\s}\s^t\s^{t'}\exp(\Xi(x,\s))}{\Theta(x)}\Bigr )\\
\label{eq10}
\ee
 
We have numerically solved the previous non linear equations for the models
(a) and (b) at different values of $M$ ranging from 2 to 15.  Similarly
as done in \cite{GoLa} we have extrapolated the different parameters $q$
and $R(t-t')$ to the $M\to\infty$ limit. We have found that a second
degree polynomial in $\frac{1}{M}$ yields very stable and good results.
In case of model (a) we found a phase transition at $T_{cl}^{(a)}=2.81\pm
.01$ while for model (b) we obtain $T_{cl}^{(b)}=2.11\pm 0.01$. The spin 
summation
method yields the thermodynamic quantities with good precision but is
inadequate to obtain the critical exponents at the transition.

\section{Monte Carlo results}

In order to characterize the quantum critical point we have done Monte
Carlo (MC) numerical simulations of class of models (a) and (b). While
model (a) displays strong Monte Carlo sampling problems (and needs a lot
of computational time) the model (b) yields the critical behavior with
modest computational effort. Note that model (a) corresponds the case
considered in references \cite{RiYo,GBHu}.  In what follows, and
otherwise stated, we will present numerical results for model (b). In
order to simulate the system described by eq.(\ref{eq3}) we consider $M$
coupled systems along the time direction with the same realization of
disorder.  To increase the speed of the computations we have considered
the case of discrete couplings $J_{ij}=\pm\frac{1}{\sqrt(N)}$ which
yields the same behavior in the large $N$ limit as in the case of a
Gaussian distribution of couplings. We have simulated two different
replicas $\lbrace\s_i^t,\tau_i^t; i=1,..,N; t=1,..,M\rbrace$ of the
system eq.(\ref{eq5}) with the same disorder realization.  The main
quantity we are interested in is the spin-spin overlap

\be
Q=\frac{1}{NM}\sum_{i=1}^N\,\sum_{t=1}^M\,\s_i^t\tau_i^t
\label{eq11}
\ee

which yields the spin-glass susceptibility,

\be
\chi_{SG}=N (\overline{\langle q^2\rangle}-\overline{\langle q\rangle}^2)
\label{eq12}
\ee

Following \cite{RiYo,GBHu} we consider the Binder parameter for different
values of $N$ and $M$. This adimensional parameter measures the
Gaussianity of the statistical fluctuations and is defined by,

\be
g=\frac{1}{2}[3-\overline{(\frac{\langle q^4\rangle}{\langle q^4\rangle^2})}]
\label{eq13}
\ee

In the vicinity of the critical point the spin-glass susceptibility 
eq.(\ref{eq12}) and the Binder parameter eq.(\ref{eq13}) 
are expected to scale with the size of the system $N$ and the temporal 
dimension $M$ in the following way,

\be
\chi_{SG}=N^p\hat{\chi} (N (T-T_c)^q,N/M^r)\\
\label{eq14}
\ee
\be
g=\hat{g}(N (T-T_c)^q,N/M^r)\\
\label{eq15}
\ee

where $\hat{\chi},\hat{g}$ are scaling functions and $p,q,r$ are
mean-field exponents related to the exponent $\nu$ and the dynamical
exponent $z$.  \footnote {This exponent z should not to be confused with
the dynamical exponent asociated with the critical-time dynamics in classical
systems.}

Now we face the problem that the finite-size scaling depends on two
variables $N,M$. As noted in \cite{RiYo} the phase transition is
signalled by the behavior of the parameter $g$ as a function of $N$ and
$M$.  For large values of $M$ the system behaves as a one-dimensional
system and for small values of $M$ the system behaves as the classical
SK model. Then the Binder parameter (\ref{eq13}) is expected to go to
zero for large and small values of of $M$. At intermediate values of $M$
a maximum for $g$ is expected. Above the critical temperature the system
becomes disordered and the value of $g$ associated to that maximum
decreases with $N$. Below $T_c$ it increases with $N$ since the system
tends to order.  At the critical point $T=T_c$ the maximum value of $g$
is constant with $N$.  According to eq.(\ref{eq14}) the scaling with $N$
of the value of $M$ corresponding to the position of maximum determines
the mean-field exponent $r$. The previous criterium yields the critical
temperature with very good precision. We find $T=2.11\pm .01$ in
agreement with the results that we obtained in the last section. Our
results for the spin-glass susceptibility $\chi$ and the Binder
parameter $g$ are shown in figures 1 and 2 at $T=2.11$. The values of
$N$ we studied cover the range $N=32-160$ with $5000$ samples in each
case. We have observed that small values of $N$ (in fact , less than
$N\simeq 50$) are affected by strong subdominant corrections to the
critical behavior. The reason is easy to understand since in the model
we are studying the maximum of $g$ is located at quite small values of
$M$ (for instance, at $N=32$ the position value of $M$ where the $g$ has
its maximum is located at a value less than $2$ which is certainly very
small).

Larger values of $N$ (we only show data for $N$ larger than $64$) allow
to extract the values of the critical exponents. The exponents $p, q, r$
can be derived as a function of $\nu$ and the dynamical exponent $z$.
These are given by, $\nu=\frac{pq}{2}$, $z\nu=\frac{q}{r}$ which yield
$\gamma=2\nu$.  The numerical results for $g$ show that the exponent
$r=\frac{1}{2}$ fits very well the scaling of the function $g$ at the
critical point. The fit of the spin glass susceptibility as a function
of the temperature in the region of scaling $M=0.42 N^{\frac{1}{2}}$ is
shown in the inset of figure 2 and is quite consistent with
$q=\frac{3}{2}, p=\frac{1}{3}$ which yields $\nu=\frac{1}{4}$ and
$\gamma=\frac{1}{2}$ as predicted within the Gaussian approximation
\cite{RS}.  Unfortunately it is difficult for us to conclude, from the
numerical data, on the exact value of the exponent $z$. Our best fit
reveals $r=\frac{1}{2}$, $z=3$ which yields $\beta=\frac{7}{8}$. But it
is very plausible that these exponents are an artifact of the
aforementioned subdominat finite $M$ corrections.  On the light of these
considerations we expect that the {\em canonical} exponents
$r=\frac{2}{3}$ , $z=4$ (which would also yield $\beta=1$) are the
correct ones.  These are the values of the exponents used to scale data
in figures $1$ and $2$. To definitely conclude on this point we should
explore laregr sizes. But this is a very difficult task due to the
long-ranged nature of the model we are studying which makes simulations
very much time consuming. It is interesting to note that the critical
value of $g$ ( $g_c=Max (g(N,M,T_c))$ ) is close to 0.056 and smaller
than the values obtained in two and three dimensions \cite{RiYo,GBHu} as
expected. As previously said we have also performed numerical
simulations of model (a) which shows a critical value of $g$ of order
$0.07$ slightly higher than that of model (b). But in this case we have
not been able to make the data for $g$ to collapse in a single universal
curve. We are suspicious that strong Monte Carlo sampling problems are
the reason for such bad results. This is presumably related to the value
of $B$ in the critical point which is higher in model (a) than in model
(b). This implies stronger anisotropic interactions in the first case.

Recently  Miller and Huse have obtained the imaginary-time correlation
function at the critical point using a theoretical analysis \cite{MiHu}.
Our mean-field exponents are in disagreement with their results.
At the critical point they obtain, 

\be
C(t)=\langle \s_i^0\s_i^t\rangle \sim t^{-\alpha}
\label{eq16}
\ee

with the value $\alpha=2$.  Figure 3 is a check of the theoretical
prediction by Miller and Huse for the imaginary-time correlation
function at the critical temperature $T=2.11$ .  Simulations have been
done for a large system $N=2272, M=20$ such that it is in the scaling
region where we expect the $g(N,M,T_c)$ takes its maximum value. We have
carefully checked that the system is in thermal equilibrium and data has
been averaged over 8 samples.  The results for the decay of the
correlation function eq.(\ref{eq6}) yields an exponent $\alpha\simeq
1.2$ consistent with the exponent $\alpha=\frac{\beta}{\nu z}$ which
ranges from $1$ to $7/6$ depending if $z=3$ or $z=4$. Note that the
decay of the imaginary-time correlation function eq.(\ref{eq16}) is
quite sensitive to how much close we are to the critical
region. Obviously, if we are not precisely in the critical region we
expect the system to be slightly more disordered and the correlation
function to decay faster. In any case the fitted value $1.2$ is an upper
limit to the true exponent $\alpha$ which we find natural to be $1$ and
then $z=4$. It is not clear to us how the predicted exponent $\alpha=2$
can fit the numerical data.

\section{Zero-temperature relaxational dynamics}

Once we have characterized the zero temperature quantum transition we
want to present some results concerning the real-time dynamical behavior
of the quantum model at zero temperature. We face the problem of
defining a reasonably real-time dynamics for a quantum system. We have
considered the simple possibility that real time Monte Carlo dynamics is
an appropiate tool to explore the slow dynamic process in the presence
of tunneling effects. In the classical case (zero transverse field) we
already know that the relaxation at zero temperature of the system stops
whenever it founds a metastable state \cite{Opper}. Because the dynamics
is non ergodic in the classical case (the system cannot jump over energy
barriers) then the system cannot reach the ground state energy. When a
transverse field is applied the system can jump over energy barriers
allowing for a new type of relaxation.  In order to investigate this
point we have considered the relaxational dynamics of the {\em true}
quantum model of eq.(\ref{eq3}) with the coefficients $A,B,C$ given in
eq.(\ref{eq3b}) at very low temperatures as a function of the transverse
field. Concretely we are interested in the behavior of the model for
large $\beta$ in the limit $M\to\infty$ with $\frac{\beta}{M}$ as much
small as possible \footnote{Note that in eq.(\ref{eq2}) the limit $\beta
\rightarrow \infty$ is performed after the limit $M \rightarrow \infty$
}.

In this limit the Hamiltonian eq.(\ref{eq3}) is strongly anisotropic,
the coefficent $A$ goes like $\frac{1}{M}$ while $B$ is much larger and
goes like $log(M)$. The total energy in eq.(\ref{eq3}) can be decomposed
in two parts plus a configuration independent constant $C$: $E=A E_J + B
E_F +C$ where $E_J$ is the sum of all interaction energies in the
different imaginary-time slices and $E_F$ is a nearest-neighbour
ferromagnetic interaction between spins in the different imaginary-time
slices. Our main quantity of interest is the relaxational behavior of
the interaction part $E_J$ as a function of time. We will show that the
dynamical evolution of the system is the same if the Monte Carlo time is
rescaled by the factor $(\frac{\beta}{M})^2$. This is a natural result
since the parameters $A$ and $B$ of the effective Hamiltonian of
eq.(\ref{eq6}) are only a function of that ratio. Note that in the limit
$M\to\infty$ the relaxation of the energy $E_J$ is extremely slow with
time (because the main contribution to the full energy in the
Hamiltonian eq.(\ref{eq3}) is due to the ferromagnetic term $E_F$).
This clarifies the apropriate regime of parameters $\beta$ and $M$ in
which the zero temperature relaxation of the model is defined. Moreover,
depending on the values of $\beta$ and $M$ one is considering, it
unambigously determines a diferent region of the real dynamical time which
is explored.

We performed two kind of experiments.  We have studied zero temperature
dynamical relaxations at a fixed transverse field. We have considered
the model at different low temperatures and different values of $M$ such
that yield nearly the same thermodynamic properties. In figure 4 we show
the relaxation of the energy $E_J$ as a function of Monte Carlo time for
different values of $M$ and $\beta$ such that the ratio
$\frac{\beta}{M}$ is small.  The simulations were performed for two
different sizes $N=320, 640$ finding the same qualitative results.  We
studied several different ratios $\frac{\beta}{M}$ ranging from $0.2$ to
$0.001$. The explored temperatures were $T=0.1, 0.02, 0.01$, deep in the
zero-temperature region, and the values of $M=50, 100, 500,
1000$. Relaxations were studied with a small transverse field
$\Gamma=0.1$ (the critical values of the transverse field is close to
$1.6$ \cite{GoLa}). In order to make the relaxation curves collapse in a
single curve we have rescaled the time by the factor
$(\frac{\beta}{M})^2$. It is interesting to observe that the energy
$E_J$ decreases to a value close to $-0.76$ which is the expected value in
the classical SK model at zero temperature at first order of replica
symmetry breaking \cite{MPV}. Note also that the energy
$E_J$ decreases with time but it can fluctuate and increase due to the
tunneling effects in the presence of the transverse field. Indeed we have
clearly appreciated this effect especially in the large-time regime.

Another interesting aspect of the quantum model we are considering
concerns its glassy properties due to tunneling effects.  The transverse
field controls the intensity of quantum fluctuations and we expect
strong hysteresis effects as the transverse field is varied. This is
shown in figure 6 where we plot the relaxation of the energy $E_J$ at
three different {\em cooling-heating} rates as a function of the
transverse field $\Gamma$ \footnote{In our case the parameter which is
varied is the transverse field and not the temperature as in real
glasses}. The cooling rate is defined by the number of Monte Carlo steps
per temperature step ($\De T$=0.05 in figure 6).  Hysteresis curves for
different values of $M$ and $\beta$ collapse in the same curve once the
cooling rate is appropiately scaled by the time factor
$(\frac{\beta}{M})^2$.  The area enclosed in the hysteresis curves
decreases as the cooling-heating rate decreases very similar to what
happens in real glasses.


\section{Conclusions}

In this work we have studied the zero temperature behavior of the
infinite-range quantum Ising spin glass in a transverse field. In
particular we have studied the critical properties at the quantum
transition point and the relaxational behavior as a function of the
transverse field.

Concerning the static properties we have studied an effective model (the
so called model (b)) which is expected to be in the same universality
class as the original quantum model eq.(\ref{eq3},\ref{eq3b}). Also this
effective model does not present strong Monte Carlo sampling problems
and gives enough confident results.  Even though our results for model
(b) show strong finite $M$ corrections for small sizes, our data is in
agreement with the mean-field quantum exponents $\nu=1/4, \beta=1,
\gamma=1/2, z=4$ \cite{RS}. Unfortunately we have not been able to
corroborate the prediction of Miller and Huse for the imaginary time
autocorrelation function eq.(\ref{eq16}) where $\al=2$. This is the
result expected for a dynamical quantum exponent $z=2$ which we
definitely rule out from the analysis of the data shown in figures 1 and
3. In particular, numerical data shown in figure 3 reveals an exponent
of $\al\simeq 1.2$ which should be a little bit lower if we are not
precisely within the scaling region. The value $\al=1$ seems us the
natural exponent compatible with our numerical results. This is an
interesting point which deserves further investigation. Unfortunately it
is very difficult to go to larger sizes since we would need much more
computing time.

Concerning the dynamical properties of the model we have investigated
the zero temperature relaxational behavior of the model. We have found
that the quantum model eq.(\ref{eq3},\ref{eq3b}) in the zero temperature
limit $\beta\to\infty$, with $\frac{\beta}{M}\to 0$ shows a universal
relaxational behavior when the Monte Carlo time is rescaled by the factor
$(\frac{\beta}{M})^2$. For a low transverse field we have observed that
the universal curve for the interaction energy $E_J$ monotonically
decreases approximately to the static value predicted in the classical
SK model at first order of replica symmetry breaking (obviously there
are small corrections due to the finite value of the
transverse field).  Because the effective model (\ref{eq3},\ref{eq3b})
mainly depends on the ratio $\frac{\beta}{M}$ we expect that similar
conclusions about the dynamical behavior of the infinite-ranged model
are also valid in the short-ranged case. We have also observed the
glassy features of the model by studying the hysteresis effects as a
function of the {\em cooling-heating} rate variation of the transverse
field. The results shown in figure 5 indicate a dynamical behavior of
the model reminiscent of that observed in real glasses.  In the presence
of a transverse field the system can jump over energy barriers due to
tunneling effects.  Then, at zero temperature, the system is not
constrained to remain forever in a metastable state. It can be
instructive to speculate if this jumping of the system over the energy
barriers corresponds to some kind of activated processes in classical
glassy models. This interesting point deserves further investigation.


\section{Acknowledgments}
We are grateful to J. Gonzalez Carmona for discussions on this and
related subjects. J. V. A. has been supported by a grant of the
Universidad Carlos III de Madrid. F.R. has been supported by the DGICYT
of Spain under grant PB92-0248 and the EC Human Capital and Mobility
Programme contract ERBCHRXCT930413.

\vfill\eject

\vfill
\newpage
{\bf Figure Captions}
\begin{itemize}

\item[Fig.~1] Binder parameter $g(N,M)$ in model (b) at $T=T_c=2.11$ for
different sizes $N=64,96,128,160$ as a function of $M/N^{\frac{2}{3}}$.

\item[Fig.~2] Spin-glass susceptibility $\chi(N,M)/N^{\frac{1}{3}}$ in
model (b) at $T=T_c=2.11$ for different sizes $N=64,96,128,160$ as a
function of $M/N^{\frac{2}{3}}$.  The inset shows the
$\chi(N,M)/N^{\frac{1}{3}}$ as a function of the temperature for $N=32,
96, 160$ for values of $N,M$ where the $g$ takes its maximum value.

\item[Fig.~3] imaginary-time correlation function in model (b) for
$N=2272$, $M=20$ in the scaling region averaged over 8 samples. The fit
is of the form $C(t)=A/t^{\al}+A/(20-t)^{\al}$ with the best fit
parameters $\al=1.2$, $A=0.3$.

\item[Fig.~4] Relaxation of the energy $E_J$ with
$\Gamma=0.1$ for different ratios $\beta/M$ as a function of the
rescaled Monte Carlo time $t^*=t (\frac{\beta}{M})^2$.

\item[Fig.~5] Hysteresis cycles of the energy $E_J$ at three different
cooling rates $r$ (dotted line $r=100$, dashed line $r=10$, continuous
line $r=1$).

\end{itemize}

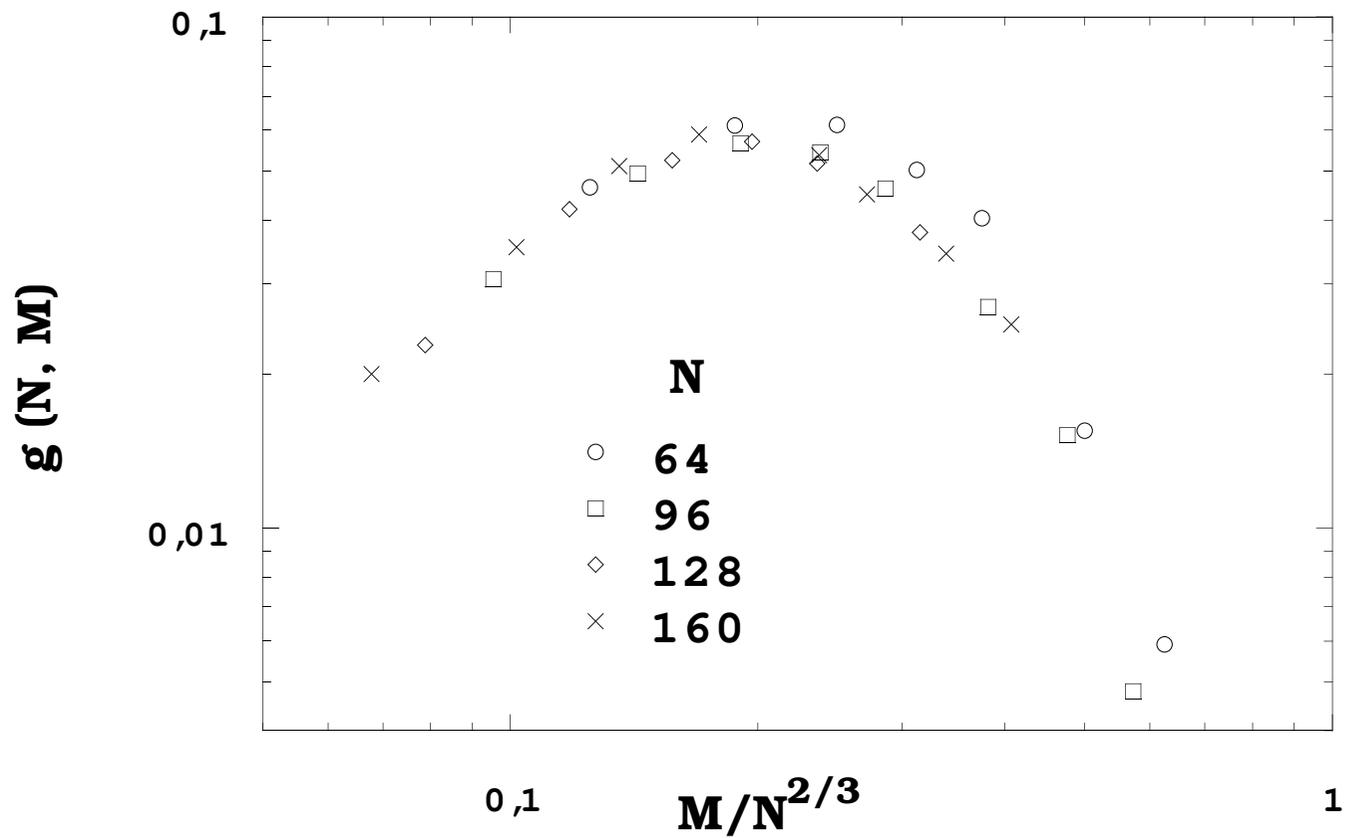
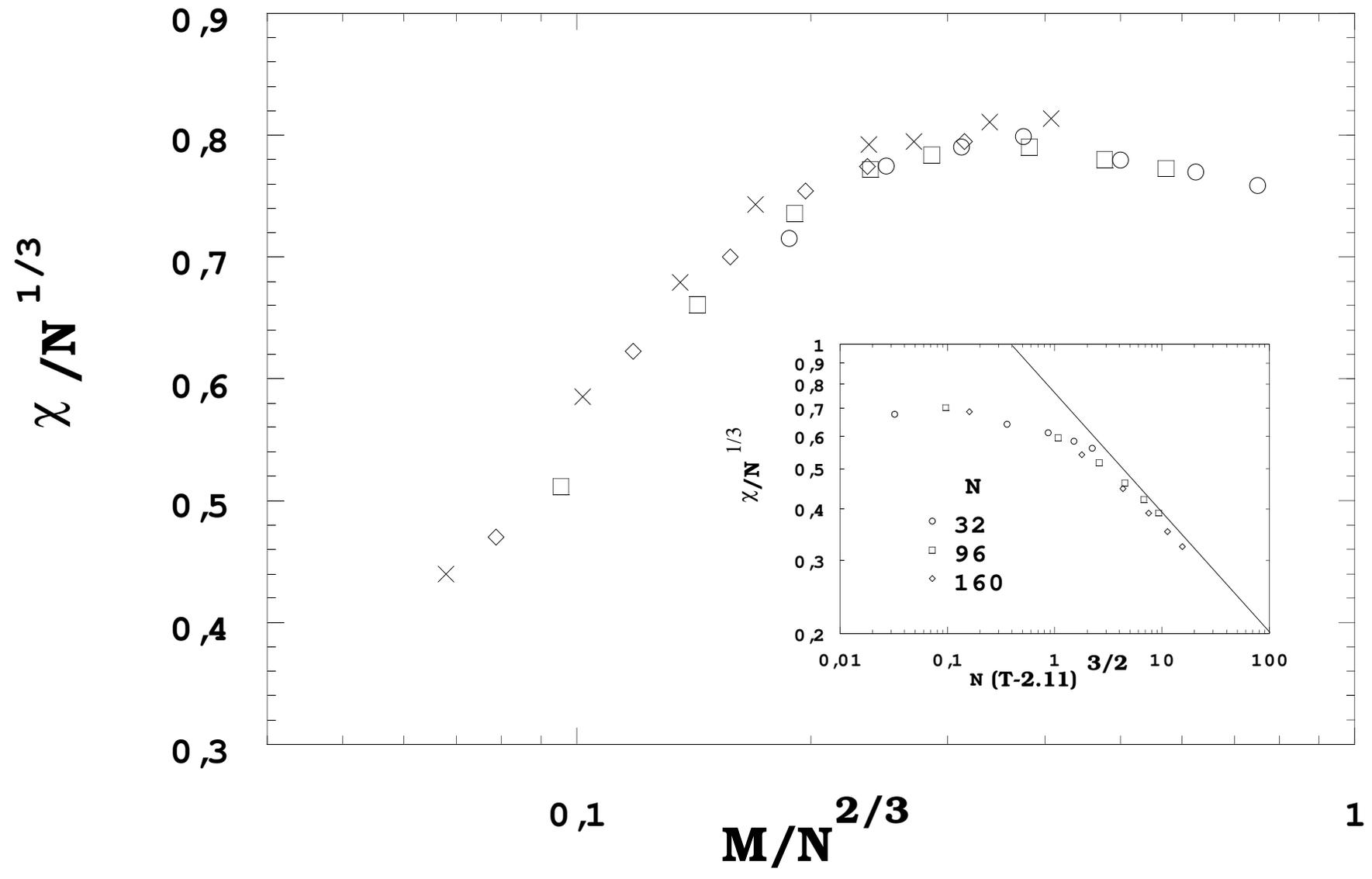
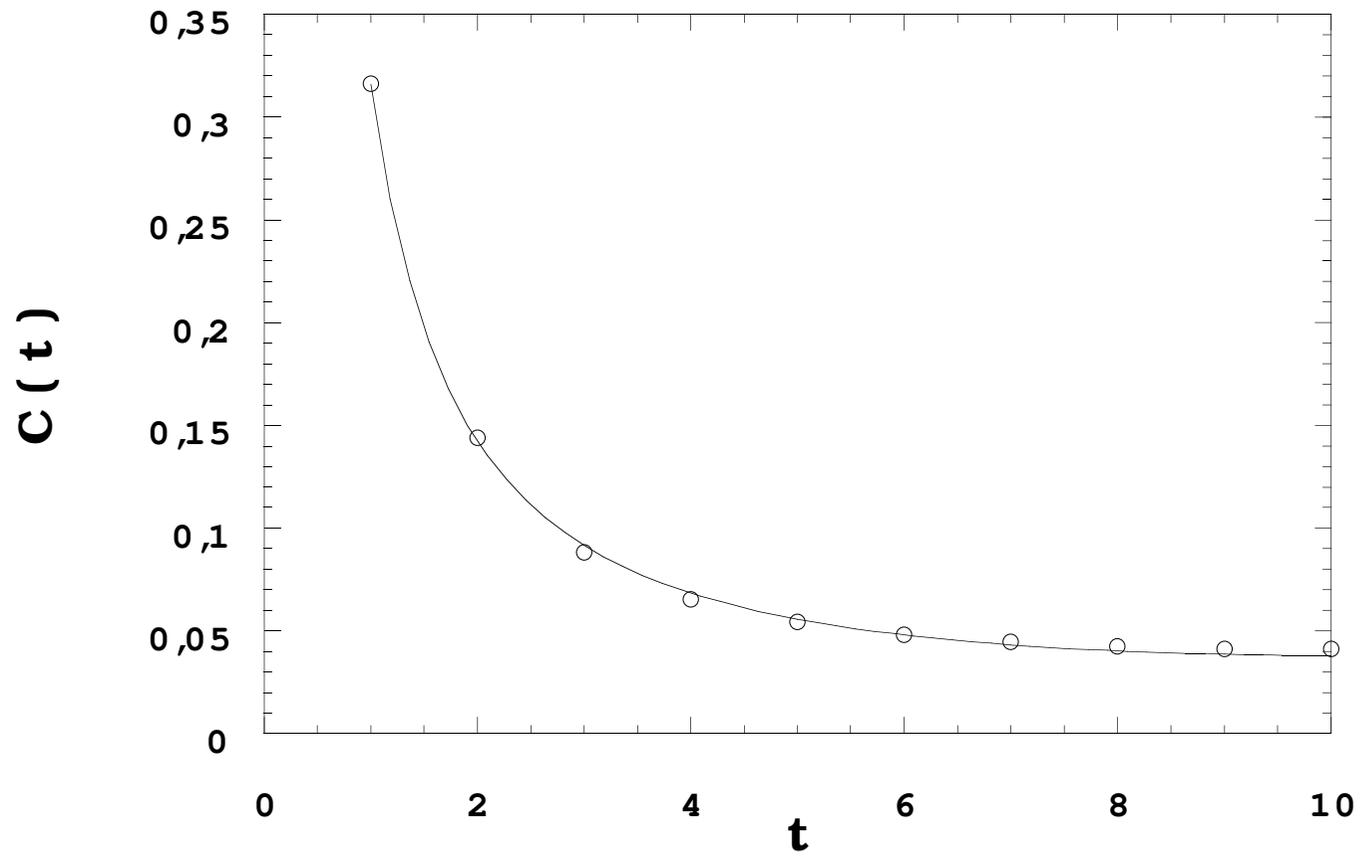
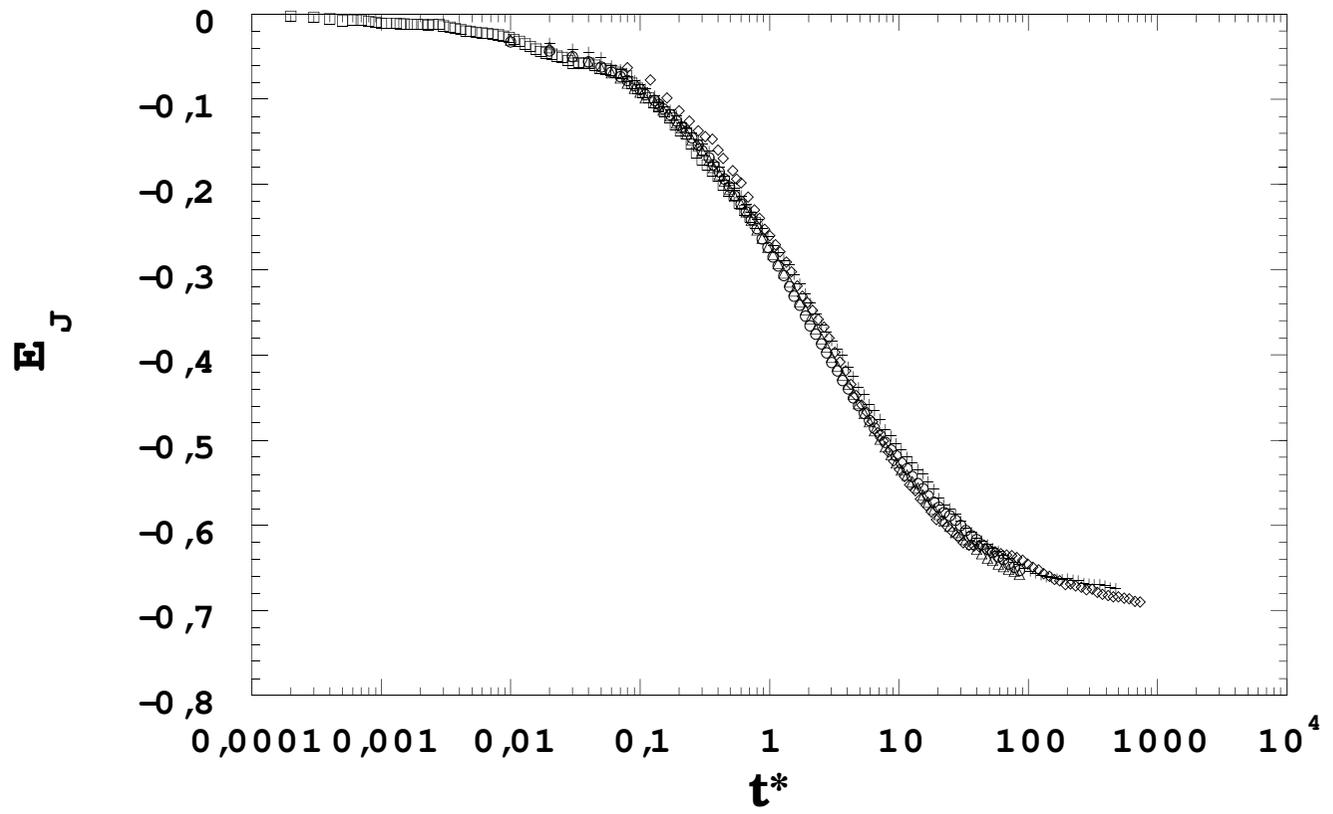
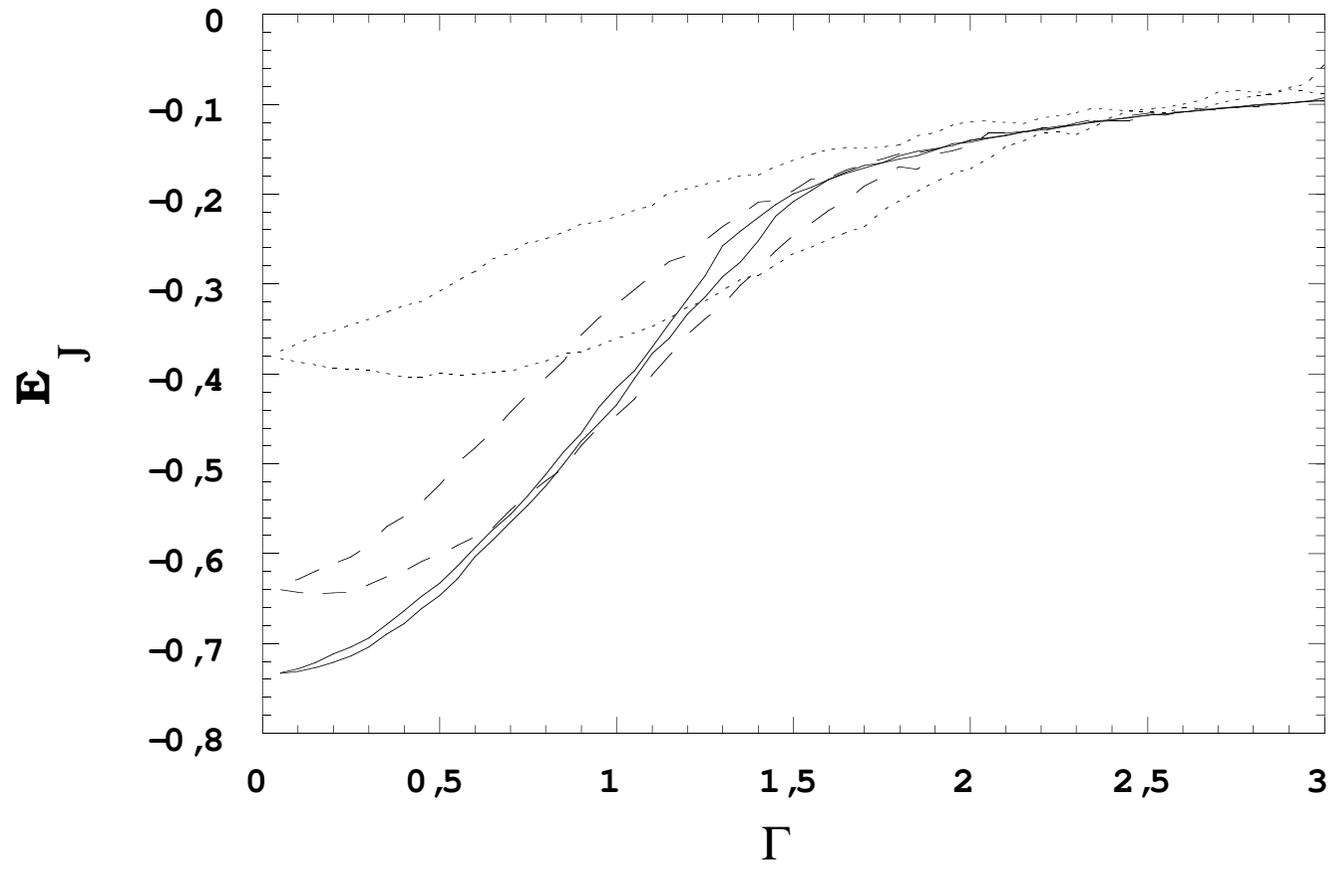
\end{document}